\documentclass[12pt,preprint]{aastex}

\shorttitle{Chemistry in a forming disk}
\shortauthors{Aikawa et al.}

\begin{document}


\title{CHEMISTRY IN A FORMING PROTOPLANETARY DISK: MAIN ACCRETION PHASE}


\author{Haruaki Yoneda}
\affil{Department of Planetology, Kobe University, Kobe 657-8501, Japan}

\author{Yusuke Tsukamoto}
\affil{Riken, 2-1 Hirosawa, Wako, Saitama, Japan}

\author{Kenji Furuya}
\affil{Center for Computational Sciences, University of Tsukuba, Japan}

\and

\author{Yuri Aikawa}
\affil{Center for Computational Sciences, University of Tsukuba, Japan}
\email{aikawa@ccs.tsukuba.ac.jp}

\begin{abstract}

We investigate the chemistry in a radiation-hydrodynamics model of star-forming core which evolves from a cold
($\sim 10$ K) prestellar core to the main accretion phase in $\sim 10^5$ yr. A rotationally-supported gravitationally unstable disk
is formed around a protostar. We extract the temporal variation of physical parameters
in $\sim 1.5 \times 10^3$ SPH particles which end up in the disk, and perform post-processing calculations of the gas-grain chemistry adopting a three-phase model.
Inside the disk, the SPH particles migrate both inward and outward. Since a significant fraction of volatiles such as CO can be
trapped in the water-dominant ice in the three-phase model, the ice mantle composition depends not only on the current position in the disk but also
on whether the dust grain has ever experienced higher temperatures than the water sublimation temperature. 
Stable molecules such as H$_2$O, CH$_4$, NH$_3$ and CH$_3$OH
are already abundant at the onset of gravitational collapse and simply sublimated as the fluid parcels migrate inside the water snow line.
On the other hand, various molecules such as carbon chains and complex organic molecules (COMs) are formed in the disk.
COMs abundance sensitively depends on the outcomes of photodissociation and diffusion rates of photofragments in bulk ice mantle.
As for S-bearing species, H$_2$S ice is abundant in the collapse phase. In the warm regions in the disk, H$_2$S is sublimated to be destroyed, 
while SO, H$_2$CS, OCS and SO$_2$ become abundant.
\end{abstract}


\keywords{astrochemistry --- star-formation --- protoplanetary disks}



\section{INTRODUCTION}

Stars are formed by gravitational collapse of dense cloud cores. Theoretically, this process is divided into two phases: early collapse phase
and main accretion phase. The former starts from the gravitational instability of a cold starless core. Efficient cooling by dust continuum
keeps the collapsing core cold. Eventually, the contraction heating overwhelms the cooling. The enhanced pressure gradient halts the collapse in the
central region to form the first hydrostatic core. Dissociation of H$_2$ triggers the collapse of the central region of the first core, which evolves to
a protostar when the dissociation and ionization of gas are completed. The initial mass of a protostar is predicted to be as small as Jovian mass
\citep{larson69, masunaga00, machida10}.
Then the protostar acquires mass mostly from the infalling envelope in the main accretion phase. Observationally, the protostellar cores, 
especially those as young as Class 0, correspond to the main accretion phase. Although these star-formation processes would be common among
high-mass and low-mass stars, we concentrate on low-mass stars, which are relevant to planetary system formation, in the present work.

Protostellar cores harbor rich chemistry \citep{vandishoeck95, ceccarelli07}. The low-mass protostellar cores with emission lines of
complex organic molecules (COMs) are called hot corinos. A single-dish line survey towards a prototypical hot corino source, IRAS 16293-2422, has
detected more than 70 species, including COMs such as CH$_3$CHO and HCOOCH$_3$ \citep{caux11}. Furthermore, recent ALMA observations
detected glycolaldehyde, methyl formate, acetic acid and ethylene glycol towards this source \citep{jorgensen12, jorgensen16}.
Glycolaldehyde and ethylene glycol are also detected towards NGC1333-IRAS2A \citep{coutens15, maury14}.
\citet{maury14} observed COMs towards NGC 1333-IRAS2A to find out that the emission lines originate from a region of radius 40-100 AU
centered on the protostar. More recently, \citet{taquet15} performed interferometric observations of NGC1333 IRAS 2A and IRAS 4A. They
also confirmed that most of the COMs emission originates in the central ($\sim 2 ''$) region of the core, and derived the abundances of COMs
to be $\sim 10^{-10}-10^{-8}$ relative to hydrogen. L1527 and IRAS 15398-3359, on the other hand, are characterized by
warm  carbon chain chemistry (WCCC) \citep{sakai08, sakai09}.  

Both COMs and carbon-chains are considered to be formed by combination of gas-phase and grain-surface reactions \citep{herbst09}.
\citet{garrod06} calculated gas-grain chemistry assuming a constant density and increasing temperature to mimic the
heating via star-formation. They found that icy radicals, which are largely produced by cosmic ray-induced photodissociation,
start to migrate on grain surfaces to form COMs at several tens of K, and that the gas-phase reactions are strongly coupled to this process
as well.
\citet{aikawa08} and \citet{aikawa12} then adopted the chemical reaction network of \citet{garrod06} to the 1D (spherical symmetric) radiation
hydrodynamics model of low-mass star formation \citep{masunaga98,masunaga00} . They investigated spatial distributions of
molecular abundances and their dependencies on the evolutionary stage of the core. They also confirmed that carbon chains are formed
at luke-warm temperatures triggered by the sublimation of methane, as suggested by \citet{sakai08}.
A recent statistical survey indicates that carbon chains and COMs indeed coexist in some protostellar cores \citep{graninger16, imai16}.

At the center of a core, a protostar (or protostars) and disk system are formed. Thus COMs and carbon chains may
eventually be incorporated into the disk. A recent ALMA observation indicates that COMs are associated with disk-forming region,
at least in the case of IRAS 16293 \citep{oya16}. Theoretical models of Class II disks, on the other hand, show that chemistry in the disk midplane
would not be efficient enough to significantly change the abundances and hydrogen isotope ratios of  stable icy molecules,
especially if the cosmic-ray ionization is prohibited by stellar winds and/or magnetic fields \citep{aikawa99,cleeves14}, although
the gas-phase chemistry is active in the disk surface and warm molecular layers \citep{dutrey97, aikawa02, qi11, rosenfeld13, henning13}.
Thus chemistry in protostellar cores and forming disks could play an important role in determining the raw material of  planetary systems.

Disk formation processes, in turn, could affect the molecular evolution in the central region of the protostellar core.
Theoretical models have predicted that COMs are efficiently formed at several tens of K. If the collapse is spherical (i.e. no rotation and thus without disk), 
gas and dust in the envelope fall onto the protostar in the free-fall timescale. Then COMs formation in the infalling material is limited
by the short timescale for them to go through the luke-warm regions ($\lesssim 10^3$ AU) of the envelope. If the gas and dust are incorporated into the disk,
COMs formation could proceed in the disk for a longer timescale. Chemistry in forming disks has been studied by several groups.
\citet{furuya12} and \citet{hincelin13} adopted the results of 3-D (magneto-) radiation hydrodynamics models \citep{tomida10,
commercon11} to calculate molecular evolution from a prestellar core to the first hydrostatic core, which is the precursor of
a protoplanetary disk. They found that the first core harbors COMs, although their abundances are yet limited due to the short lifetime ($\lesssim$ a few $10^3$ yr)
of the first core and more compact luke-warm regions than those in the protostellar stage.
\citet{visser09, visser11} constructed a two-dimensional semi-analytical model of disk formation, starting from the singular isothermal
sphere. Although the model proceeds to the protostar phase (i.e. beyond the first core), their grain-surface chemistry is limited to rather simple
reactions such as hydrogenation. Evolution of complex organic molecules in the semi-analytical model was recently investigated by
\citet{drozdovskaya16}. They showed that COMs abundance can be comparable to that of CH$_3$OH in the disk midplane, when all the
envelope gas has accreted onto the disk.

The present work is an update from \citet{aikawa08} and \citet{furuya12}; we investigate the molecular evolution
from a molecular cloud core to a forming disk around a protostar (i.e. beyond the first core) with an updated chemical model.
We adopt the
3-D radiation hydrodynamics (RHD) simulation of \citet{tsukamoto15}. Full radiation hydrodynamics calculation after the first core is very time
consuming, because of the extremely short time step required to calculate the structure and evolution of a protostar \citep{tomida13}.
\citet{tsukamoto15} circumvent this problem by introducing a sink particle at the protostar position to successfully calculate the
evolution up to $10^4$ yr after the formation of the protostar.  We solve the rate equations of gas-grain chemistry along the stream lines of fluid parcels,
which are traced in the RHD \citep[e.g.][]{aikawa13}. Since the fluid parcels spend longer time in warm disk regions, we can expect more significant chemical
evolution than in the first core stage.
While the chemical network model used in this work is basically the same as  that used in \citet{aikawa08} and \citet{furuya12},
we adopt the three-phase model rather than the two-phase model.

The rest of this paper is organized as follows. The physical and chemical models are described in \S 2. The results of our fiducial model are
presented in \S 3. In \S4, we compare our fiducial model with the two-phase model, and investigate dependences of our results on some uncertain chemical parameters. We briefly
compare our results with recent models and observations of forming disks. Finally, we summarize our conclusions in \S 5.

\section{MODELS}
\subsection{Disk Formation}
We adopt our disk formation model from a 3D radiation hydrodynamical simulation by \cite{tsukamoto15}.
The simulation starts from an isothermal, uniform and rigidly rotating molecular cloud core with  mass of 1 $M_{\odot}$ and temperature of
$T=10$ K. \citet{tsukamoto15} investigated models with various initial angular velocity and central density
to derive the criterion for disk fragmentation. We adopt the model in which a relatively massive ($>0.2 M_{\odot}$) disk is formed;
angular velocity of rigid rotation, density, and core radius at the initial conditions are $\Omega=7.61\times 10^{-14}$ s$^{-1}$,
$\rho=6.91\times 10^{-19}$ g cm$^{-3}$, and $r=5.8\times 10^3$ AU, respectively.
The gravitation collapse of the rotating core is calculated by the three-dimensional smoothed particle radiation hydrodynamics (SPRHD) code, which can treat the
heating by the gas dynamics and radiative heating/cooling. We used the sink particle and does not include radiative feedback from the sink particle.
Thus, it may underestimate the temperature in the disk and in falling envelope after the protostar is formed.

\begin{figure}
\epsscale{.80}
\plotone{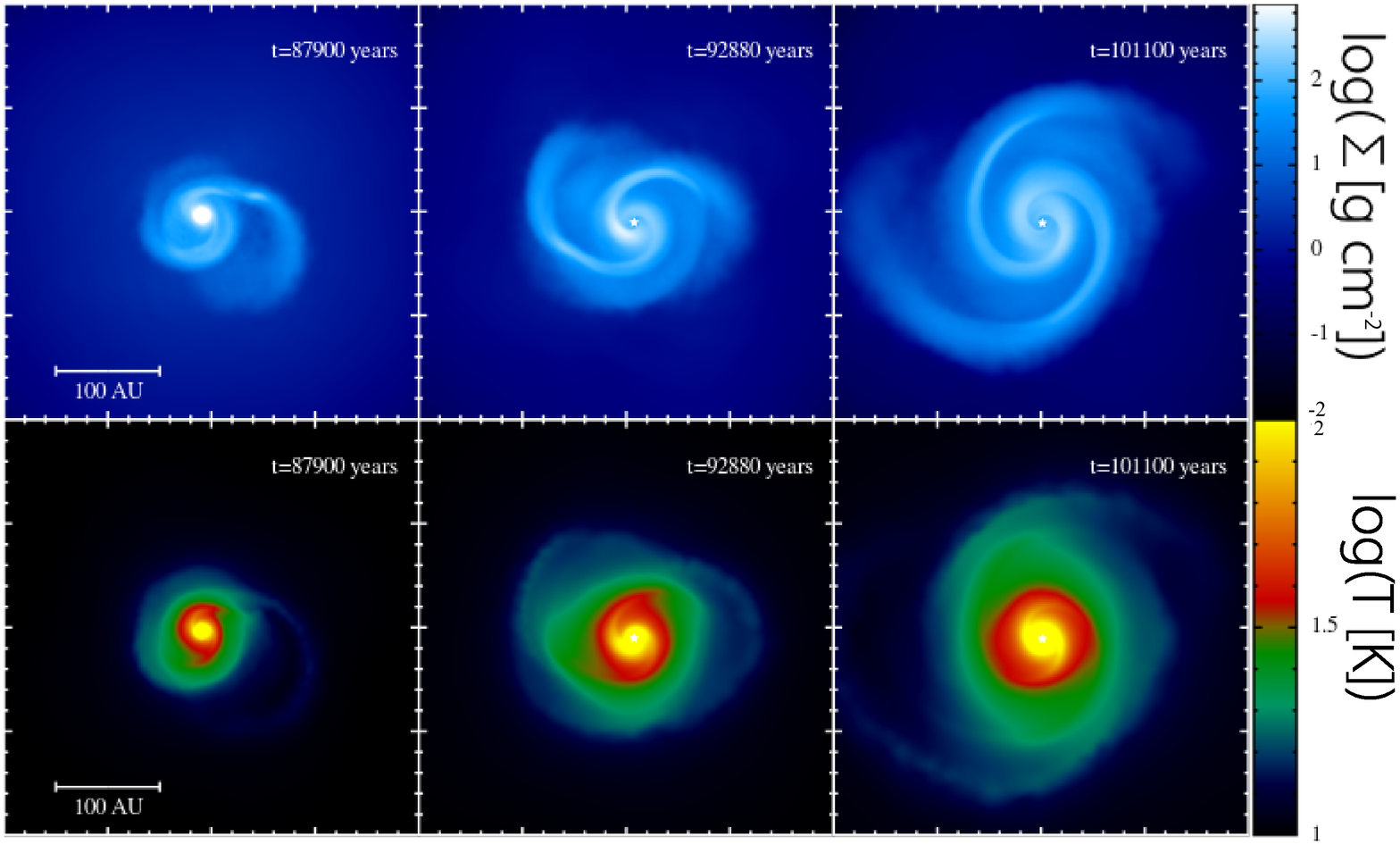}
\caption{Distribution of the column density (a) and density-weighted temperature (b) viewed face-on at assorted time steps in the disk formation model by \citet{tsukamoto15}.
\label{disk_formation}}
\end{figure}

Figure \ref{disk_formation} shows the distribution of column density and density-weighted gas temperature
($\int \rho T dz/\int \rho dz$)
at assorted time steps at the core center on $x-y$ plane, which is perpendicular to the rotation axis.
At the time of $t=t_{\rm SC}=9.046 \times 10^4$ yr, the central density reaches $\sim 4\times 10^{-8}$ g cm$^{-3}$.
Then the gas inside the radius of 2 AU is replaced by a sink particle, which represents a protostar.
The simulation continues up to $t=t_{\rm f}=1.011\times 10^5$ yr, when the masses of the protostar
and the disk are 0.107 $M_{\odot}$ and 0.221 $M_{\odot}$, respectively. Here, the disk is defined as the rotationally-supported region;
i.e. the centrifugal force is larger than a half of the gravitation force towards the center. 
Since the disk is more massive than the protostar,
the rotation curve is shallower ($\Omega \propto r^{-1.1}$) than the Keplerian law ($\Omega \propto r^{-1.5}$), and the spiral arms are
formed by the gravitational instability. 

Among the $5.2\times 10^5$ SPH particles used in the RHD simulation, we extract about $1.5\times 10^3$ particles, which reside near
the disk midplane at $t_{\rm f}$, to calculate the chemical rate equations along their trajectories, i.e. taking into account the temporal variation
of the gas density, temperature and UV radiation. The visual extinction $A_{\rm v}$ is used to calculate the attenuation of the interstellar 
UV radiation \citep[e.g.][]{furuya12}.
We evaluate the visual extinction of each SPH particle by calculating the column densities of interstellar matter from the particle position
to the boundaries of the computational domain along the $x$, $y$ and $z$ axes in both the positive and negative directions.The minimum column 
density among the 6 directions is converted to the visual extinction via the formula
\begin{equation}
A_{\rm v}=\frac{f_{\rm H}\Sigma}{\mu m_{\rm H}}\times 5.34\times 10^{-22} {\rm cm}^2,
\end{equation}
where $f_{\rm H}$ is the mean number of hydrogen nuclei per species (1.67), $\mu$ is the mean molecular weight (2.3) and $m_{\rm H}$
is the mass of a hydrogen atom \citep{bohlin78, cardelli89}. Since we adopt the SPH particles which end up in the disk midplane,
$A_{\rm v}$ is mostly high ($\gtrsim 3$ mag) so that the photodesorption and photodissociation by interstellar radiation field are not important
in our model. UV radiation from the protostar is not taken into account (see discussion in \S 4.4).
The cosmic-ray induced UV radiation, however, is ubiquitous unless the column density is much higher than
96 g cm$^{-2}$ (see \S 2.2).

\begin{figure}
\epsscale{.80}
\plotone{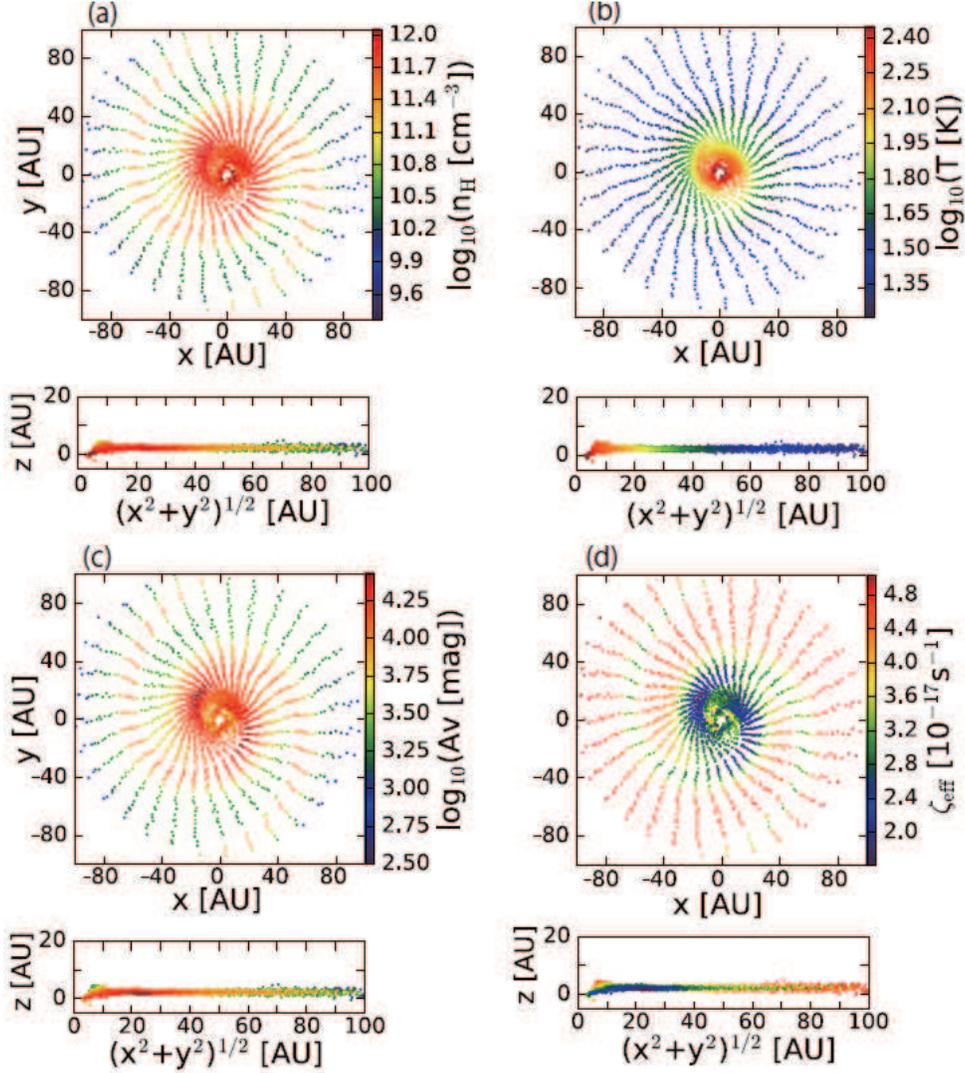}
\caption{Distribution of extracted SPH particles color coded by the number density of hydrogen nuclei (a), temperature (b), the visual extinction (c),
and the ionization rate (d) at the final time step of our simulation.
The upper panels show the projection onto the $x-y$ plane (i.e. midplane)  and the lower panels show the distribution in
the radius-$z$ plane.  The $z$-axis corresponds to the rotational axis. 
\label{disk_sph}}
\end{figure}

Figure \ref{disk_sph} shows the distribution of extracted SPH particles color-coded according to their gas density $n_{\rm H}$ (panel a),
temperature (b), visual extinction (c) and cosmic-ray ionization rate (d). While the density distribution shows the spiral structure due to the 
gravitational instability, the temperature distribution is rather axisymmetric; the heating by the spiral arm is weak and transient,
and does not significantly contribute to the temperature distribution \citep{tsukamoto15}.


\subsection{Chemical Model}

Our chemical network model is basically the same as in \citet{furuya12}. The gas-phase reactions and their rates are adopted from
\citet{garrod06} at low temperatures ($T\le 100$ K), while they are adopted from \citet{harada10} at higher temperatures.
Reaction rate coefficients are updated following recent references; major updates include nitrogen chemistry (Wakelam et al. 2013, arXiv:1310.4350)
and branching ratios of dissociative recombination of COM ions \citep[e.g.][]{geppert06,garrod08}.
Three-body reactions and collisional dissociation reactions are included in our model; they could be
effective in the central regions with high density ($\gtrsim 10^{12}$ cm$^{-3}$) and high temperature ($\gtrsim$ several
100 K) \citep{furuya12}. The ionization rate is set to be
\begin{equation}
\zeta_{\rm eff} = 5\times 10^{-17} \exp \left\{-\frac{\Sigma}{96 {\rm g cm}^{-2} }\right\}+1\times 10^{-18} {\rm s}^{-1}.
\end{equation}
The first term is the ionization rate by cosmic rays and the second term represents the ionization by decay of radio-active nuclei.

The treatment of grain-surface reactions and interactions between the gas phase and grain surfaces is the same as in \citet{furuya15}. 
The gas/dust mass ratio is set to be 100, and a uniform grain size of 0.1 $\mu$m is assumed.
Grain-surface reactions and desorption energies of atoms and molecules are adopted from \citet{garrod06}.
We consider the Langumuire-Hinshelwood mechanism for grain surface reactions; i.e. the atoms and molecules on grain surfaces migrate
via thermal hopping and react with each other when they meet.
The diffusion energy barrier ($E_{\rm b}$) is set to be half of the desorption energy ($E_{\rm d}$) of each species.
We adopt the modified rate (method A) of \citet{garrod08}, which takes into account the competition between surface
processes (e.g. accretion versus diffusion) to improve the accuracy of the original rate equation method when the number of
reactants is small on a grain surface. We take into account the competition between the reaction and diffusion out of a
binding site on grain surfaces. The competition effectively enhances the probability to overcome the activation barrier, when the
diffusion timescale is larger than the timescale to overcome the activation barrier. It should be noted that the activation barrier
could be overcome by tunneling, while the diffusion is assumed to be thermal in our model.
We also take into account three non-thermal desorption processes: photodesorption, stochastic heating
by cosmic-rays, and reactive desorption \citep[e.g.][]{oberg09a,hase93,garrod07}. The yield of photodesorption per incident photon is set to be $10^{-3}$.
The efficiency of reactive desorption is $10^{-3}-10^{-2}$ per reaction for most species; it is lower for products with more vibrational modes.

In dense cold cores, ice mantles become thicker than a monolayer.
Then the question is if the diffusion (hopping) rate is the same on the surface and in deeper layers. It is intuitive to expect that
diffusion would be less efficient in deeper layers, if the ice mantle is dense. In fact, laboratory experiments show that hydrogen atoms
adsorbed onto ice mantle can penetrate and react with only the several surface monolayers \citep{fuchs09,ioppolo10}.
There are, however, other experiments that suggest diffusion of molecules in bulk ice. \citet{oberg09c}, for example, found that the water
ice mixture containing either CO$_2$ or CO, segregate through surface diffusion and bulk diffusion. The effective diffusion rate in the deep
layers would depend on the structure of the ice mantle, e.g. cracks and porosity.
Thus we calculate two models: the two-phase model and the three-phase model. The former corresponds to the case with very
efficient bulk diffusion, while the latter corresponds to the other extreme, i.e. no bulk diffusion.
We adopt the three-phase model as our fiducial model and compare the two models in discussion (\S4.1).
 
The two-phase model consists of gas phase and ice phase (i.e. grain surface species). All icy molecules and atoms are assumed to
migrate via thermal hopping and to be chemically active.  In the three-phase model, ice phase is divided to surface layers and inner bulk \citep{hase93,garrod11}.
Only the surface layers, which is set to be 4 monolayers in the present work, are subject to diffusion, reaction and desorption. 
Both in the two-phase and three-phase models, we assume that only the top 4 monolayers are subject to photodissociation, and calculate their rates accordingly \citep{furuya15}. In other words, although the UV radiation can penetrate into the deeper layers of ice mantle, we assume that photofragments recombine immediately to reform their mother molecules rather than migrate and form new molecules.
The dependence of molecular evolution on this assumption is discussed in \S 4.2.

We assume the so-called low-metal elemental abundance \citep[Table 2 of][]{furuya12}. The molecular abundances at the onset of gravitational collapse
are set by calculating a molecular evolution with the three-phase model for $1\times 10^6$ yr under cloud core conditions: the temperature of $T=10$ K,
the density of $n_{\rm H}=2\times 10^4$ cm$^{-3}$, the visual extinction $A_{\rm v}=10$ mag, and the cosmic-ray ionization rate
$\zeta_{\rm CR}=5\times 10^{-17}$ s$^{-1}$ \citep{dalgarno06}.

\section{RESULTS}
\begin{figure}
\epsscale{1.0}
\plotone{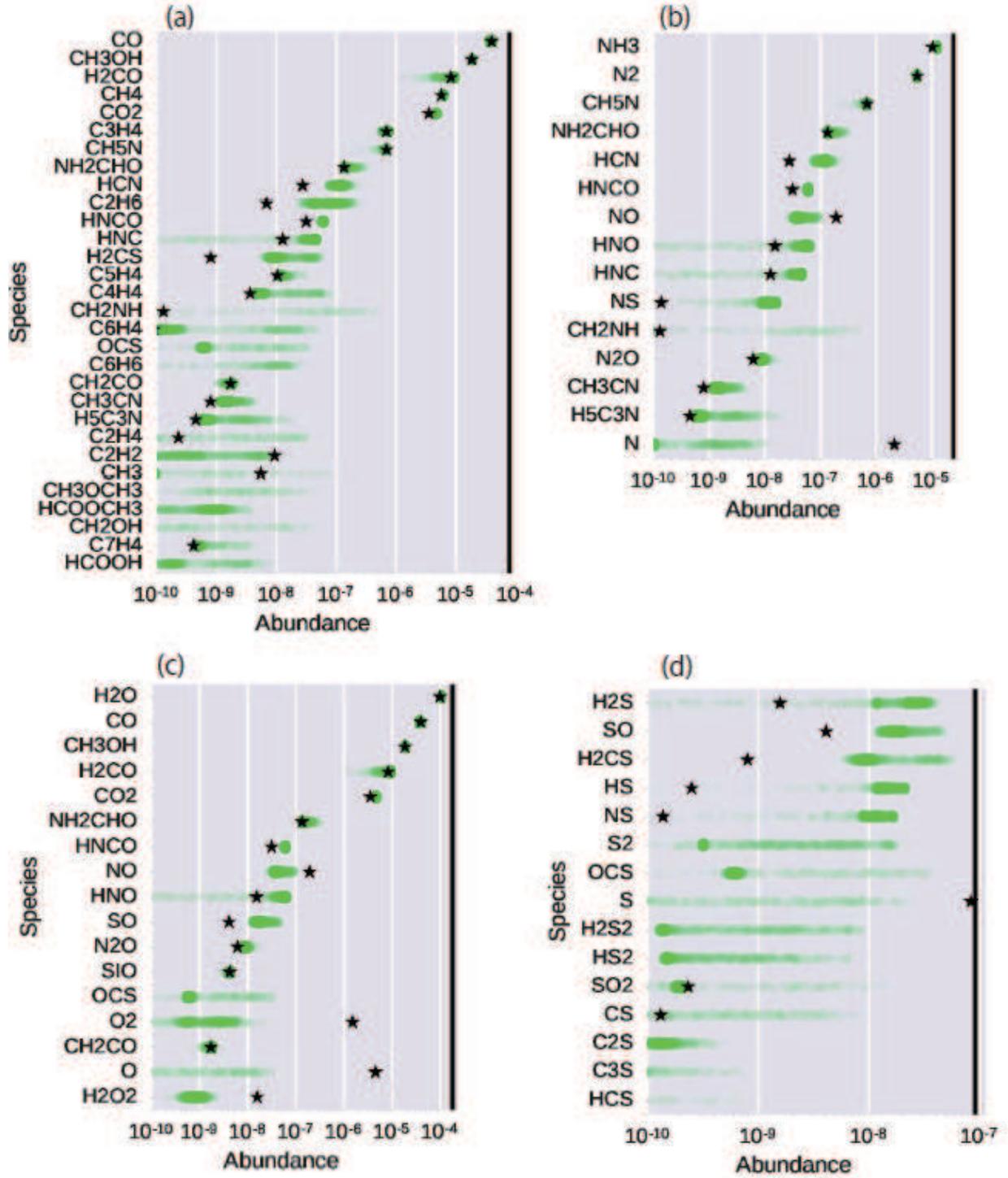}
\caption{
 Frequency distribution of the total (gas and ice) abundances of the most abundant molecules ($n({\rm i})/n_{\rm H}\gtrsim
10^{-10}$) at the final time step. The thickness of the green circle is proportional to the number of SPH particles with the
corresponding molecular abundances. The star symbols depict the initial abundance. 
\label{top30}}
\end{figure}

Figure \ref{top30} shows the frequency distribution of the total (gas and ice) abundances of the most abundant molecules at $t=t_{\rm f}$ for (a) C-bearing,
(b) N-bearing, (c) O-bearing and (d) S-bearing species in the $\sim 1.5\times 10^3$ SPH particles. The color indicates the number of SPH particles
with the specific molecular abundance; it is thick green, if the number of SPH particles is 20 or larger. The initial abundances are denoted by the star symbols. 
While the major stable molecules, such as CO and H$_2$O,
keep their initial abundances, less abundant molecules with the initial abundance of $\lesssim 10^{-5}$ tend to evolve significantly during the simulation.
In order to understand when and how these molecules are processed, we pick up one SPH particle and describe its physical and chemical 
evolution in the following subsection.

\subsection{Molecular Evolution in a SPH particle}

\begin{figure}
\epsscale{1.0}
\plotone{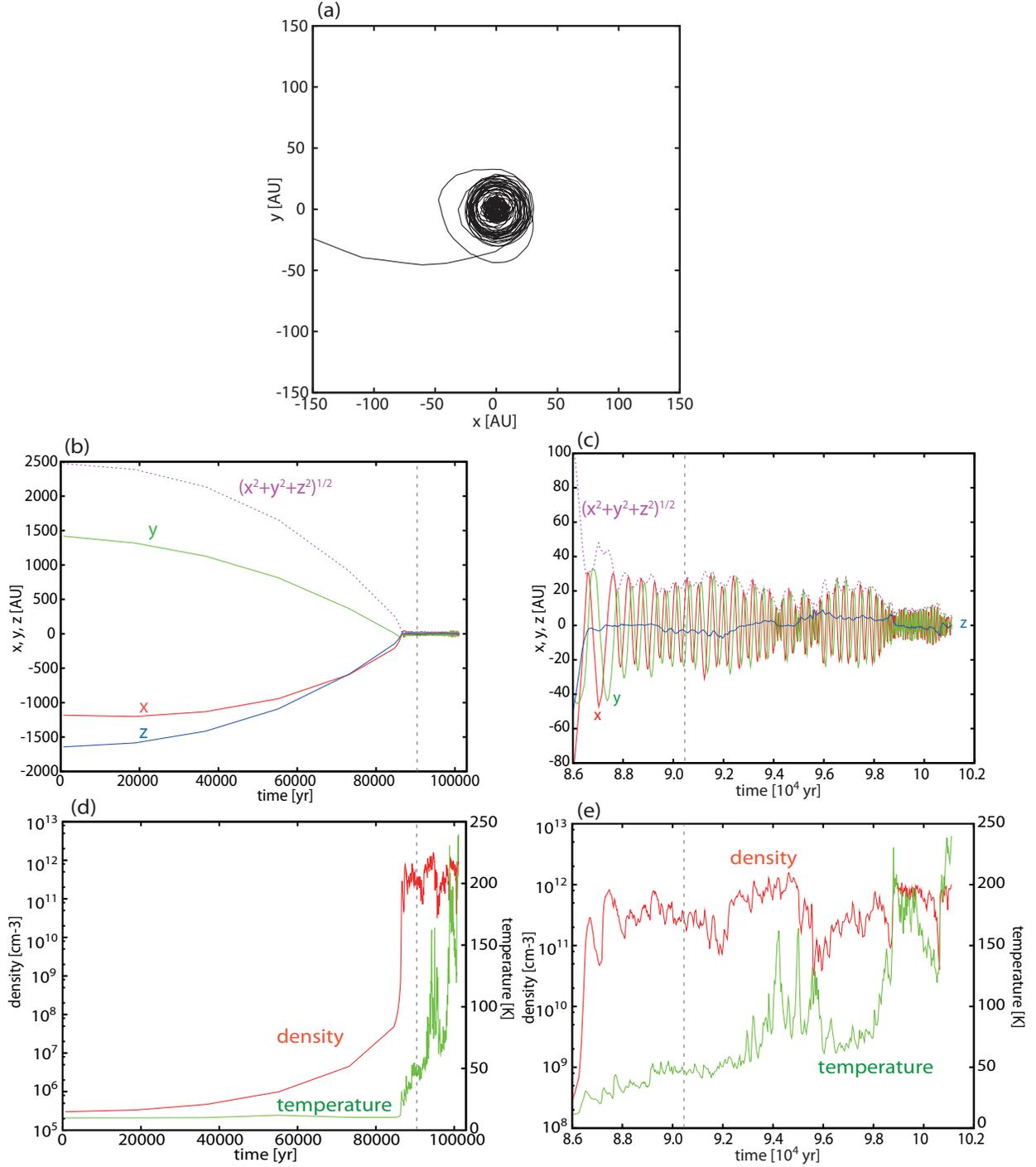}
\caption{Trajectory (a-c) of a SPH particle and the temporal variation of the physical parameters (d-e) in the SPH particle. The protostar is formed at $t=9.046\times 10^4$ yr,
which is indicated by the gray dashed line.
\label{traj1948}}
\end{figure}

Figure \ref{traj1948} shows the trajectory (a-c) and the temporal variation of density and temperature (d-e) of a SPH particle,
which was initially at the radius of $2.5\times 10^3$ AU. 
At $\sim 8.6 \times 10^4$ yr, the SPH particle falls onto the forming disk, and starts to rotate around the first core.
Figure \ref{traj1948} (c) and (e) shows the trajectory and physical parameters of the SPH particle after entering the disk.
While the disk is rotationally supported, the trajectory
is fluctuated by the dynamical effect such as the spiral arm, which results in the temporal variation of  the temperature and density.
At $t=t_{\rm SC}$ (i.e. $9.046\times 10^4$ yr), the central region of the first core collapses to form the protostar,
while the rotationally-supported outer regions become the circumstellar disk. 

\subsubsection{infalling phase}
Temporal variation of molecular abundances in the SPH particle is shown in Figure \ref{evol1948}.
The left column shows the evolution in the infalling phase, $t \le 8.65 \times 10^4$ yr, in which the density gradually increases,
while the temperature remains almost constant $\sim 10$ K. As the density rises, gaseous molecules are depleted onto grains.
The major icy molecules, CO, CH$_3$OH, H$_2$CO, and CH$_4$ among C-bearing species, H$_2$O among O-bearing species,
and NH$_3$  among N-bearing  species,  are abundantly formed before the onset of collapse.

Among S-bearing species, atomic sulfur is the most abundant at the
onset of gravitational collapse. It is converted to HS and H$_2$S by the following gas-phase reactions,
\begin{eqnarray}
{\rm OH} + {\rm S} &\rightarrow& {\rm HS} + {\rm O},\\
{\rm HCO}^+ + {\rm HS} &\rightarrow& {\rm H}_2{\rm S}^+ + {\rm CO},\\
{\rm H}_2{\rm S}^+ + {\rm S} &\rightarrow& {\rm S}^+ + {\rm H}_2{\rm S},
\end{eqnarray}
and a series of hydrogenation on grain surfaces
\begin{eqnarray}
{\rm H_{ice}} + {\rm S_{ice}} &\rightarrow& {\rm HS_{ice}},\\
{\rm H_{ice}} + {\rm HS_{ice}} &\rightarrow& {\rm H_2S_{ice}}.
\end{eqnarray}
On the grain surface, H$_2$S can be converted back to HS via H$_{\rm ice}$ + H$_2$S$_{\rm ice}$ $\rightarrow$ H$_{\rm 2 ice}$ + HS$_{\rm ice}$.
Although this reaction has an
activation barrier of 860 K, it proceeds as efficiently as the hydrogenation of HS, due to the competition between the reaction and H atom diffusion.
The abundances of HS and H$_2$S ices are thus similar \citep{furuya15}.
SO is formed in the gas phase via S + OH $\rightarrow$ SO + H, and on grain surfaces via
HS$_{\rm ice}$ + O$_{\rm ice}$ $\rightarrow$ SO$_{\rm ice}$ + H$_{\rm ice}$.

\subsubsection{forming disk}
The right column of Figure \ref{evol1948} shows the molecular evolution after the SPH particle entered the forming disk.
Both density and temperature fluctuate in a very short timescale, $\sim$ a few hundred years. Molecular abundances are
more sensitive to temperature than to density. For example, at the time of $\sim 9.4\times 10^4$ yr, the temperature rises from
$\sim 50 $ K to $\sim 150$ K, as the SPH particle migrates inwards. Then the major icy molecules, such as CO, CH$_4$, and H$_2$S,
sublimate to the gas phase. Although the sublimation temperature of these molecules are much lower than 150 K, they have been
trapped in water-dominant ice mantle, which is characteristic of the three-phase model \citep[e.g.][]{vasyunin13}, and is often observed in sublimation of mixed ice
in the laboratory \citep[e.g.][]{collings04}.

CO, CH$_4$, H$_2$O, NH$_3$, and CH$_3$OH  remain abundant in the gas phase after sublimation in our model, partly
because we follow the evolution only for $\sim 10^4$ yrs after the sublimation. Astrochemical models of
hot cores \citep[e.g.][]{charnley92, millar98} and T Tauri disks \citep{nomura09} show that
CH$_4$, NH$_3$ and CH$_3$OH are destroyed by the gas-phase reactions in $10^4-10^5$ yr after sublimation. This
timescale is not sensitive to the gas density, because the main reactants are ions, as long as the ionization rate is 
$\sim 10^{-17}$ s$^{-1}$ and the temperature is $\lesssim$ a few 100 K.
On the other hand, H$_2$S, which also forms abundantly in the envelope, is destroyed
by the reaction with atomic hydrogen (see below)  \citep{millar98}.

The abundances of complex organic molecules such as HCOOH\footnote{Although HCOOH is made of only five atoms, we count it as COMs, because
it is characteristic of hot corinos.}, HCOOCH$_3$ and CH$_3$OCH$_3$ increase in the disk.
They can form by both gas-phase and grain-surface reactions.
In the gas-phase,  HCOOH forms mainly by OH + H$_2$CO (see discussion in \S 4.3).
HCOOCH$_3$ forms by H$_3$CO$^+$ +  H$_2$CO
$\rightarrow$ H$_2$COHOCH$_2^+$ and subsequent dissociative recombination. CH$_3$OCH$_3$ forms by
H$_3$CO$^+$ + CH$_4 \rightarrow$ CH$_3$OCH$_4^+$ and subsequent recombination. This protonated ion can also
form by CH$_3^+$ + CH$_3$OH, once the SPH particle migrates inside the  CH$_3$OH snowline. Due to the
high gas density ($\sim 10^{11}$ cm$^{-3}$), these gas-phase products are promptly adsorbed onto grains at $T \lesssim 150$ K.
On grain surfaces, HCOOH forms by OH$_{\rm ice}$ + HCO$_{\rm ice}$, and HCOOCH$_3$ forms by HCO$_{\rm ice}$ + CH$_2$OH$_{\rm ice}$, while
CH$_3$OCH$_3$ forms by CH$_{\rm 3 ice}$ + CH$_2$OH$_{\rm ice}$. It should be noted that CH$_2$OH and CH$_3$O are not discriminated in our model.

Carbon chains also form in the disk. Once the gaseous methane becomes relatively abundant ($\sim 5\times 10^{-7}$) at $8.8\times 10^4$ yr,
it reacts with C$^+$ to produce C$_2$H$_3^+$ or C$_2$H$_2^+$. The latter reacts with H$_2$ to form C$_2$H$_4^+$.
Then C$_2$H$_4^+$ and C$_2$H$_3^+$ dissociatively recombine with electron on grain surfaces to form acetylene, C$_2$H$_2$. C$_2$H$_6$ forms
by successive hydrogenation of C$_2$H$_2$ on grain surfaces. In the later stage ($\gtrsim 9.8\times 10^4$ yr), C$_2$H$_2$ also
forms by CH$_2$ + CH$_2$ $\rightarrow$ C$_2$H$_2$ + H + H in the gas phase. C$_3$H$_2$ and C$_4$H, which
are thought to characterize the WCCC sources such as L1527, are not abundant in the forming disk due to the high density and high temperature \citep{aikawa12}.
C$_3$H$_2$ is protonated, and  C$_4$H is transformed to C$_4$H$_2$ via the gas-phase reaction with H$_2$.

While the abundances of H$_2$S and HS are almost the same in the envelope, the partial sublimation of HS starts earlier than that of
H$_2$S due to higher volatility. Then HS reacts with atomic H to form atomic S, or reacts with atomic S to form S$_2$. 
The sublimation of major S-bearing ices at $\sim 9.4\times 10^4$ yr triggers active gas-phase chemistry. H$_2$S is destroyed by 
H$_2$S + H $\rightarrow$ HS + H$_2$; although this reaction has an activation barrier of 1350 K, the warm temperature ($> 70$ K) helps to overcome
the barrier. Then SO and H$_2$CS form by O + S$_2$ and S + CH$_3$, respectively \citep{charnley97}. CS forms by CH$_2$ + S.
After the water ice sublimates into the gas phase together with the entrapped minor ices ($t\gtrsim 9.8\times 10^4$ yr),
SO$_2$ and OCS form by OH + SO, and S + CO, respectively. Since the proton affinity of OCS is higher than that of CO but lower than that of H$_2$O,
destruction of OCS via protonation, i.e. OCS + HCO$^+$ $\rightarrow$ HOCS$^+$ + CO, becomes inefficient, after the water sublimation.

It should be noted that  at warm temperatures ($\gtrsim$ a few 10 K), H$_2$ formation on ice mantle surfaces is inefficient. Then H atom,
which is the major reactant of HS and H$_2$S, becomes as abundant as $10^{-7}-10^{-6}$ relative to hydrogen (i.e. $n$(H)$\sim 10^5-10^6$ cm$^{-3}$).
At $T>100$ K, we assume the efficient H$_2$ formation via chemisorption site  \citep{cazaux04}, which lowers the H atom
abundance temporally. Duration time of such high temperature is, however, rather limited, and the H$_2$S abundance significantly decreases before the fluid parcel
reaches the innermost hot region of the disk.


\subsection{Cloud Origin or Disk Origin}

In the astrochemical studies of protoplanetary disks, it is often argued how much fraction of disk material inherits the chemical composition of ISM, and how much fraction is chemically
reset in the disk. Based on our simulation, here we discuss which molecule originates in the cloud (i.e. inherit from ISM) and which forms in the young disk (i.e. reset).
Figure \ref{top30} shows that chemical composition in the forming disk is quite different from our initial condition, which is set by calculating the
molecular evolution for $1\times 10^6$ yr under the cloud core conditions. In the previous subsection, however, we have seen that the molecular evolution proceeds
not only in the disk, but also in the cold infalling phase. For example, the abundance of H$_2$S increases, while that of C$_2$H$_2$  decreases by orders
of magnitude in the infalling phase. In the context of inherit or reset, we thus divide the major molecules into two groups according to
whether they are formed before entering the disk (i.e. before the onset of collapse and in the infalling envelope) or in the forming disk
(Table \ref{tab_origin}). We call the former group as ``Cloud'' for brevity. Each group is further divided into
two groups, based on whether or not they are destroyed in inner regions of the disk. 

Figure \ref{dist_faceon} shows the distribution of representative species at the final timestep.
CO originates in a cloud before the inset of collapse, and sum of its abundance in the gas phase and ice mantle is almost constant. H$_2$S is abundant in
the outer disk, since it is efficiently formed in the collapse phase. It is then destroyed in the gas phase after sublimation at
$\lesssim$ a few tens of AU. HCOOH and SO$_2$ form in the inner disk, and thus have a centrally peaked distribution.


\begin{table}
\begin{center}
\caption{Formation site of molecules\label{tab_origin}}
\begin{tabular}{lll}
\tableline\tableline
formation site & species & destroyed\\
\tableline
Cloud  & CO, CH$_4$, CH$_3$OH, H$_2$O, NH$_3$, CO$_2$, H$_2$CO & HS, H$_2$S \\
Disk & HCOOCH$_3$, HCOOH, C$_2$H$_6$, C$_2$H$_2$ & S$_2$, CS\\
              & CH$_3$OCH$_3$, H$_2$CS, SO$_2$, OCS& \\
\tableline
\end{tabular}
\end{center}
\end{table}

Our assignment of formation site of molecules is qualitatively similar to the result of \citet{hincelin13}, who investigated the
molecular evolution from the collapsing core to the first core. They suggested that CH$_3$OH and H$_2$S originate in the parental
cloud, while HCOOH and OCS are formed in the warm region close to the first core. But there are also significant differences between
our results. Firstly,  we include a chain reaction of CO$_2$ formation triggered by OH formation on CO ice mantle \citep{garrod11}, which makes
CO$_2$ much more abundant than in \citet{hincelin13}, although our value is yet lower than the observed CO$_2$ ice abundance in molecular clouds
\citep{bergin05}. Secondly, many molecules, such as SO and COMs,
are more abundantly produced in our model than in \citet{hincelin13}. The longer timescale and more extended region of warm temperature
in our model allow the active chemical evolution. For the same reason, the present model gives higher abundances of COMs
than our previous work which focused on the first core \citep{furuya12}.

It should be noted that our model might underestimate the chemical evolution in the envelope, since our model does not take into account
the envelope heating by protostellar irradiation. In the model of  \citet{visser11}, for example, warm temperature ($\gtrsim$ a few tens of K) regions
in the envelope extend to several 100 AU. The infalling gas and dust experience such warm temperature for a few $10^3$ yrs, during which some fraction
of COMs could be formed \citep[see also][]{aikawa12}. Our model, however, corresponds to earlier stage than that of \citet{visser11}, who
followed the longer
term evolution ($\sim 2\times 10^5$ yr after the protostellar birth). In our model, many SPH particles
are incorporated to the rotationally-supported disk before the protostar is formed. It indicates that the assignment of formation site of molecules, whether
they are mainly formed in the disk or collapsing envelope, depends on the evolutionary stage of the core.
 
\subsection{Radial Distribution}

The upper panel of Figure \ref{r_CO} shows temperature and number density of hydrogen nuclei of $\sim 1.5 \times 10^3$ SPH particles
at $t_{\rm f}$ as functions of radius.
The plotted particles
all exist at $|z| \le 0.2 R$, where $z$ is distance from the midplane and $R$ is distance from the rotation axis. Since the disk is
nonaxisymmetric with the spiral structures (Figure \ref{disk_sph}), the density and temperature have a range of values at each radius.
The CO abundance (Figure \ref{r_CO} lower panel) also shows a significant scatter especially at $R\ge 70$ AU and at 10 AU 
$\lesssim R\lesssim$ a few tens of AU.
The scatter at the outer radius is mainly caused by the variation of physical conditions; when the CO abundance in the SPH particles
is plotted as a function of temperature, the scatter becomes smaller. The scatter at the inner radii, on the other hand, is caused by radial
migration of the SPH particles. As we have seen in Figure \ref{traj1948}, the SPH particles migrate both inwards and outwards after entering the
disk. Since the dominant component of ice mantle is water, and since we adopt the three-phase model, only a fraction of CO ice is sublimated
at its own sublimation temperature $\sim 20$ K. At the sublimation temperature of H$_2$O, on the other hand, the whole mantle evaporates.
When a SPH particle looses its ice mantle and migrates out to $R\sim $ a few tens of AU, CO does not freeze out due to warm
temperatures of $T\gtrsim 20$ K.
Thus at  10 AU $\lesssim R\lesssim$ a few tens of AU, the SPH particles with high CO ice abundance ($\sim 3 \times 10^{-5}$) and low CO gas abundance coexist
with those having high CO gas abundance and low CO ice abundance.
This has three implications. Firstly our simulation clearly shows that the ice mantle composition keeps ``memory'' of the dust trajectory, so that dust
particles at similar radii could have significantly different mantle composition. Secondly, the radial abundance variation over the snow line is
smoothed by the outward migration of gas. Finally, our simulation shows that mixing of gaseous composition could be important, which is not taken
into account in the present work and should be investigated in future work.

The solid lines in the lower panel of Figure \ref{r_CO} show azimuthally averaged molecular abundances as a function of radius.
We constructed a circular coordinate with 532 radial grids and 100 azimuthal grids, and  interpolated the density and molecular abundances
at SPH particle positions to obtain the values at each grid point. Then we calculated the azimuthally averaged abundances, i.e. total number
of the molecule divided by total number of hydrogen nuclei, at each radius. The averaged abundances show spiky features especially
at low abundance range, reflecting the original scatter of the molecular abundances.

Figure \ref{radial_dist_abun} shows azimuthally averaged abundances of assorted species. The spiky features, i.e. the variations of the 
averaged abundance over a short radial distance, are caused by the scattered molecular abundances in radial bins, as we have seen for
CO (Figure \ref{r_CO}). Transition of icy 
molecules to gaseous species occur mostly at $R\sim 10$ AU, where water ice sublimates, although very volatile species such as CO and
N$_2$ have relatively high gas-phase abundances even outside the water snow line.


\section{DISCUSSION}

\subsection{Comparison with the Two-phase Model}

In the three-phase model, we assumed no diffusion in the inert bulk layer of ice mantle, which results in a significant ($\sim 85 \%$) entrapment of
CO in water ice mantle. In reality, however, molecules in the bulk ice mantle could diffuse slowly. \citet{fayolle11} found in the laboratory experiments
of H$_2$O/CO$_2$/CO ice mixture that the fraction of CO trapped in water ice varies from 4\% to 96 \%, depending on the ice mantle thickness and
ice composition. While they constructed a detailed rate equation model with swapping of ice mantle species to reproduce the laboratory experiments,
an implementation of their formulation with extrapolations to various ice species is out of the scope of the present work.
In order to investigate the dependence of molecular abundances on the bulk ice diffusion, we calculate another extreme case, the two-phase model,
in which molecules in the whole ice mantle can migrate thermally.


Azimuthally averaged molecular abundances for the two-phase model are presented in Figure \ref{radial_dist_2ph}.
An apparent difference between the three-phase and two-phase models is that, in the two-phase model, icy molecules are sublimated
at their own sublimation temperatures. Abundances at the innermost radii, which correspond to the final abundances
in the temporal variation, are similar between two-phase and three-phase models for many molecules, although there are some
exceptions.

In the two-phase model, CH$_3$OH is slightly more abundant than CO, since the whole CO ice is subject to hydrogenation,
i.e. the conversion to CH$_3$OH. Among COMs, HCOOCH$_3$ is more abundant in the two-phase model than in the three-phase model.
HCOOCH$_3$ is formed by H$_3$CO$^+$ + H$_2$CO $\rightarrow$ H$_2$COHOCH$_2^+$ and subsequent recombination in the gas phase,
and  radical-radical association of HCO$_{\rm ice}$ + CH$_2$OH$_{\rm ice}$ on grain surfaces both in the two-phase and three-phase models.
Sublimation of  H$_2$CO at $R\sim 30$ AU enhances the gas-phase formation of HCOOCH$_3$ in the two-phase model.
It also enhances the production of gaseous HCO, which is adsorbed onto grains to form HCOOCH$_3$.

C$_2$H$_2$ and C$_3$H$_2$ are sublimated at $\sim 30$ AU, where they are promptly destroyed by gas-phase reactions.
In later stage, $t\sim 9.8\times 10^4$ yr, the C$_2$H$_2$ and C$_3$H$_2$ abundances are determined by the gas-phase reactions
to be similar to those in the three-phase model.

Major N- and O-bearing species in Figure \ref{radial_dist_2ph} are formed mostly in the cloud before the onset of collapse and collapse phase,
and are simply sublimated in the disk. Their abundances in the innermost radii are similar between the two-phase and three-phase models,
as they are mostly determined by the initial abundances.

H$_2$S ice is the dominant S-bearing species outside $R \sim 20$ AU, and is converted to H$_2$CS and SO in the inner radius, as in the three-phase
model. OCS in the two-phase model is slightly less abundant than in the three-phase model, reflecting the lower CO abundance, since OCS is mainly
formed by CO + S. S$_2$ is efficiently formed via HS + S at $R\sim 40$ AU, where HS is sublimated. Then CS is formed by C + S$_2$.

\subsection{Dependences of COM Abundances on Ice Photodissociation Rate}

The previous subsection showed that COM abundances are higher in the two-phase model than in the three-phase model. The enhancement is,
however, only one order of magnitude or so. The formation of COMs is actually limited by the formation of radicals via photodissociation of icy molecule,
which is restricted to the top 4 monolayers in our model, assuming that the photofragments in deeper layers efficiently recombine rather than migrate and
form new molecules. In this subsection, we investigate the dependence of COM abundances on this assumption by letting the whole ice mantle to be
subject to photodissociation and subsequent chemical reactions in the two-phase model. The COMs formation is also expected to be enhanced
if the ratio of diffusion energy barrier to desorption energy barrier is smaller \citep{walsh14}. We thus recalculate the temporal variation of molecular
abundances in the SPH particle described in \S 3.1 with three different settings (Table \ref{models}). 
In model A, the ratio $E_{\rm b}/E_{\rm d}$ is set to be 0.3, and the photodissociation of ice mantle is effective only
in the top 4 monolayers. In model B and C, the whole ice mantle is subject to the photodissociation. The  ratio $E_{\rm b}/E_{\rm d}$ is 0.3 in Model B, while it is
0.5 in model C.

\setcounter{table}{0}
\begin{table}
\begin{center}
\caption{Model settings and COMs abundance}
\begin{tabular}{lccc}
\tableline\tableline
Model & phase & $E_{\rm b}/E_{\rm d}$ & photodiss.\tablenotemark{a} \\
\tableline
A         &  two &   0.3 &  4  \\
B         &  two &  0.3  &  all  \\
C        &  two & 0.5   & all  \\
\tableline
\end{tabular}
\label{models}
\tablenotetext{a}{Number of ice mantle layers which are effectively subject to the photodissociation.}
\end{center}
\end{table}

The top panels in Figure \ref{COMs_test} show the result of Model A.
Comparing the final COMs abundance with that in Figure \ref{radial_dist_2ph}, we can see that the COM abundances are enhanced by only a factor of a few.

In Model B (the middle panels in Figure \ref{COMs_test}), the abundances of HCOOCH$_3$ and CH$_3$OCH$_3$ are enhanced by
more than an oder of magnitude compared with those in Model A. Formation of these COMs are
efficient even in the cold ($\sim 10$ K) envelope, since the thermal diffusion barrier is set to be small especially for small (e.g. two-atom or three-atom) radicals.
In Model C (the bottom panels), COMs formation in the cold envelope is inefficient due to the high $E_{\rm b}$ compared to Model B, while
the COM abundances in the disk are higher than in Model A.  We conclude that COM abundances are more sensitive to the efficiency of ice mantle photodissociation
than on $E_{\rm b}/E_{\rm d}$. If the ice mantle photodissociation is efficient and $E_{\rm b}/E_{\rm d}$ is low, COMs can be abundantly formed even in low
temperatures.

\subsection{Uncertainties of Rate Coefficients for COMs formation}

In our model, the abundance of NH$_2$CHO is as high as $\sim 10^{-7}$ in the initial condition and remains high throughout
the simulation (Figure \ref{top30}). Its abundance estimated towards hot corino objects, on the other hand, is $\sim 10^{-11}-10^{-9}$ \citep{lopez15}.
In our model NH$_2$CHO is formed mainly by association of CH$_2$OH$_{\rm ice}$ + N$_{\rm ice}$ on grain surfaces.
This formation pathway is more effective
compared with our previous work, since we updated the photodissociation branching ratio of CH$_3$OH ice following \citet{oberg09b}, who suggested that
CH$_2$OH + H is the major branch based on the laboratory experiment.
The association reaction, however, may not lead to NH$_2$CHO, since the re-arrangement of chemical bonds is required.

HCOOH forms mainly by OH + H$_2$CO in the gas phase, and OH$_{\rm ice}$ + HCO$_{\rm ice}$ on grain surfaces in our model. The rate coefficient of the former
gas-phase reaction is set to be $2.01\times 10^{-13}$ cm$^3$ s$^{-1}$ in our model. \citet{xu06}, however, found that the association of OH + H$_2$CO
has an activation barrier of 2156 K. We may have overestimated the formation rate of HCOOH in the gas phase.

In order to check the effect of these reactions in our model, we recalculate the molecular evolution in the SPH particle described in \S 3.1, setting the initial NH$_2$CHO ice abundance to be zero and turning off the reactions CH$_2$OH$_{\rm ice}$ + N$_{\rm ice}$ $\rightarrow$ NH$_2$CHO$_{\rm ice}$ on the grain surface and
OH + H$_2$CO $\rightarrow$ HCOOH + H in the gas phase. The HCOOH abundance in the final timestep is reduced to $7\times 10^{-11}$,
but NH$_2$CHO is as abundant as $1\times 10^{-7}$ relative to hydrogen nuclei. The latter is mainly formed by the gas-phase reaction of  NH$_2$ + H$_2$CO, the rate coefficient of which is adopted from the ab-initio calculation by \citet{barone15}.

It should also be noted that several groups proposed new gas-phase formation pathways of COMs, which are not included in
our model \citep{vasyunin13b, balucani15, taquet16}. Although these new reactions were introduced to account for gaseous COMs 
observed in prestellar cores \citep{bacmann12, vastel14}, they could be effective in the warm regions of forming disk as well.


\subsection{Comparison with the Previous Models and Observations}
Chemistry in gravitationally unstable disk is previously investigate by \citet{ilee11} and \citet{evans15}. They showed that icy molecules are desorbed
to the gas phase by the shock heating at the spiral arms. Our model shows a similar feature, although the enhancement of temperature
and gas-phase molecular abundances at the spiral arms is less prominent than in \citet{evans15} (\S 2.1).
Comparing Figure \ref{disk_sph} (b) and  Figure \ref{dist_faceon}
we can see that the gas-phase molecules are abundant in the warm temperature ($T\gtrsim 100$ K) region, which is tilted by the spiral arms. Since the gas density
is higher by a factor of a few at the spiral arms than the inter-arm region, the spiral structure would become more apparent in the distribution of gaseous molecular
column densities than in the abundance distribution shown in Figure \ref{dist_faceon}. \citet{evans15} also showed that the spiral arms are well traced by CN and that
CO, SO, OCS, NH$_3$, H$_2$O and H$_2$S are significantly affected only by the adsorption and desorption. In our model, on the other hand, CN is not abundant
due to its high chemical reactivity and lower sublimation temperature ($\sim 40$ K) than assumed in \citet{evans15}. While the total (gas and ice)
abundances of CO, NH$_3$, and H$_2$O remain almost constant in our model, S-bearing species are subject to active chemistry in the gas-phase. It is not clear
why the S-chemistry is different between our model and \cite{evans15}. The reactions of S-bearing species described in \S 3.1.2 are included
in UMIST95, which is adopted in \citet{evans15}. Possible causes could be a short duration (2047 yr) of the chemical calculation and initial abundances
in \cite{evans15}. Their initial abundance is set referring to cometary ice abundance, rather than calculating the molecular evolution in the molecular clouds and/or
collapsing core; SO and OCS are relatively abundant ($\sim 10^{-6}$), while H atom and O atom are absent.

\citet{drozdovskaya16} investigated COMs formation in the semi-analytical disk formation model of \citet{visser09}. 
They found that COM abundances can be comparable to that of CH$_3$OH in the disk midplane, when all the envelope gas has accreted onto the disk.
Although there are many differences between our chemical and physical models, the major differences would be twofold. Firstly, they investigate the disk composition at the end of the envelope accretion, considering the stellar UV radiation and the outflow cavity.
The infalling material is thus significantly irradiated by stellar UV, which triggers the COMs formation by
producing radicals in ice mantle. It indicates that the molecular abundance in the envelope and disk would change as the envelope mass decrease \citep{furuya16}.
Secondly, their ice mantle chemistry is similar to our Model B in section \S 4.2;  which is the most optimistic model for COMs formation among our models.

While various COMs have been observed in hot corino sources, derivation of their abundance relative to hydrogen is not
straightforward, because non-spherical small-scale ($\lesssim 100$ AU) structures, together with the line opacities and
excitation conditions, need to be considered \citep[e.g.][]{visser13}. \citet{taquet15} observed NGC 1333-IRAS 2A and -IRAS 4A
with the Plateau de Bure interferometer and detected multiple lines of COMs coming mostly from the central compact
region ($\lesssim 0.2-0.36$ arcsec). The COMs abundances derived by \citet{taquet15} are $4\times 10^{-9}-1\times 10^{-8}$
for CH$_3$OCH$_3$, $(1-2)\times 10^{-8}$ for HCOOCH$_3$ and $4\times 10^{-7}-1\times10^{-6}$ for CH$_3$OH. 
While the abundance of CH$_3$OCH$_3$ in our fiducial model is consistent with that derived from the observation, our model
overestimates the CH$_3$OH abundance. The HCOOCH$_3$ abundance of \citet{taquet15} is better reproduced in Model A
(Figure \ref{COMs_test}) than in our fiducial model.



Observations of CO towards protostellar cores indicate that gaseous CO increases inside the radius of $T\sim 20$ K, i.e. the CO snow line, but that
the abundance of sublimated CO is lower than its canonical value ($\sim 10^{-4}$) \citep[e.g.][]{fuente12, anderl16}.
\citet{favre13} also found that gaseous CO is depleted in the warm layer of the class II disk, TW Hya.
Entrapment of CO in water ice, as observed in our three-phase model, could be an explanation for the low gaseous CO abundance at $T\gtrsim 20$ K,
although there are other possibilities such as the chemical conversion of CO to less volatile species \citep{bergin14,furuya14}. 

\citet{oya16} analyzed ALMA Cycle-1 data of IRAS 16293-2422 and found that CH$_3$OH and HCOOCH$_3$ are concentrated around the inner part of the infalling-rotating envelope.
They also found that H$_2$CS emission originates both in the envelope and the disk inside the centrifugal barrier of $40-60$ AU. 
OCS, on the other hand, mainly traces the infalling-rotating envelope, although its existence in the disk
component cannot be ruled out.
It should be noted that our model at the final timestep is still younger and colder than
IRAS 16293-2422. CH$_3$OH emission towards IRAS 16293-2422 indicates that the inner part of the infalling-rotating envelope is heated to
$\gtrsim 100$ K, either by protostellar irradiation or accretion shock. We thus compare our model results around H$_2$O snow line with the
observed chemical structure around the centrifugal barrier of IRAS 16293-2422.
CH$_3$OH and HCOOCH$_3$ are naturally sublimated to be abundant in the gas phase inside the H$_2$O snow line in our model.
H$_2$CS is abundant both in the envelope (Figure \ref{evol1948} left column) and inside the H$_2$O snow line. OCS exists in the 
envelope, as observed, but is also abundant inside the H$_2$O snow  line, where HCO$^+$ is depleted.

\section{SUMMARY}

We investigated molecular evolution from a dense molecular cloud core to a forming disk in the main accretion phase. Our findings are as follows.

\begin{enumerate}
\item{Major stable molecules such as CO, CH$_4$, CH$_3$OH, H$_2$O, NH$_3$ and CO$_2$ are abundantly formed in the cloud before the onset of collapse
and in collapsing phase, while COMs are mainly formed in the forming disk. Carbon chains such as C$_2$H$_6$ and C$_2$H$_2$ increase
in the warm regions in the disk, as well.}

\item{Among S-bearing species, H$_2$S is abundant in the collapse phase,
but it is destroyed in the gas phase upon sublimation in the forming disk. SO, H$_2$CS, OCS and SO$_2$ become abundant in warm 
regions in the disk.}

\item{Comparisons between the present work and other recent models indicate that the molecular evolution in the infalling envelope becomes more
significant as the temperature rises and density (i.e. the UV attenuation) decreases in the envelope.}

\item{In the three-phase model, only a fraction of volatile species, such as CO, is sublimated at their own sublimation temperatures, and
the remaining fraction is trapped in water ice mantle in the warmer regions.
They are fully desorbed to the gas phase at the water sublimation temperature. In the two-phase model, on the other hand,
icy molecules are sublimated at their own sublimation temperatures. Nevertheless, the gas-phase abundances at the inner most radii are rather similar between
the three-phase and two-phase models for many molecules.
}


\item{The fluid parcels (i.e. the SPH particles) migrate both inward and outward in the disk. In the three-phase model,
ice-coated dust grains with various ice composition, e.g.
with and without CO ice, co-exist outside the water snow line.}

\item{COM abundances sensitively depend on the outcomes of photodissociation and diffusion of photofragments in the bulk ice mantle.
Their abundances are significantly enhanced, if photofragments can migrate to form new molecules rather than simply recombine to
reform its mother molecule,  inside the deep ice mantle. COM abundances also increase, when a lower ratio of diffusion barrier to desorption barrier
is assumed. In the model with low diffusion barrier, COMs are formed even in the cold prestellar conditions.}

\end{enumerate}

\acknowledgments

This work was supported by JSPS KAKENHI Grant Numbers 23540266 and 16H00931.
\appendix

\clearpage



\begin{figure}
\epsscale{1.0}
\plotone{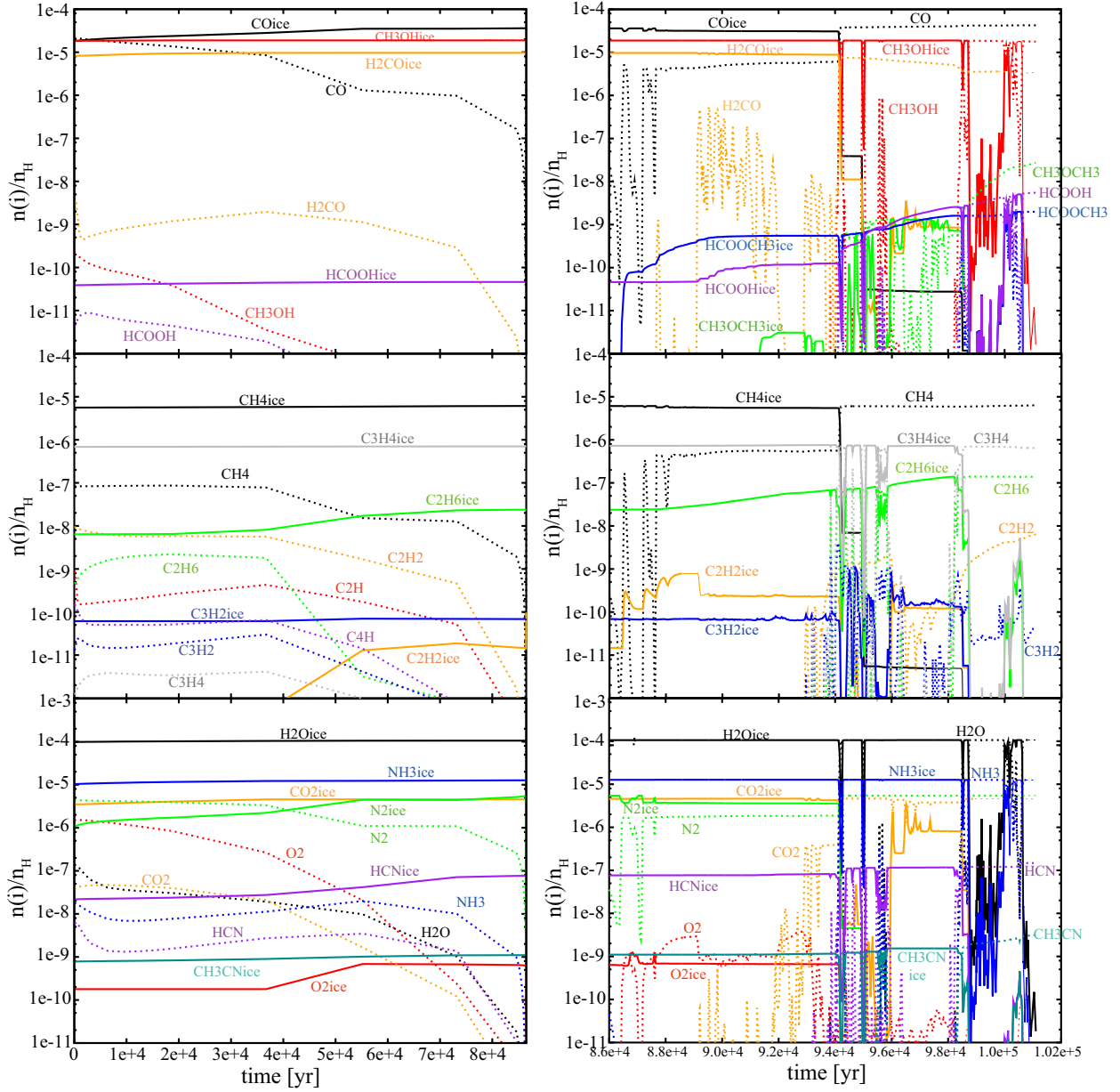}
\caption{Temporal variation of molecular abundances in the SPH particle along the trajectory shown in Figure \ref{traj1948}.
The left column shows the infalling phase, i.e. $t \le 8.65 \times 10^4$ yr, while the right column shows the molecular evolution
in the forming disk. From top to bottom, the panels show CO and complex organic molecules (COMs), hydrocarbons, 
O-bearing and N-bearing major molecules, and S-bearing molecules.
\label{evol1948}}
\end{figure}

\setcounter{figure}{4}
\begin{figure}
\epsscale{1.0}
\plotone{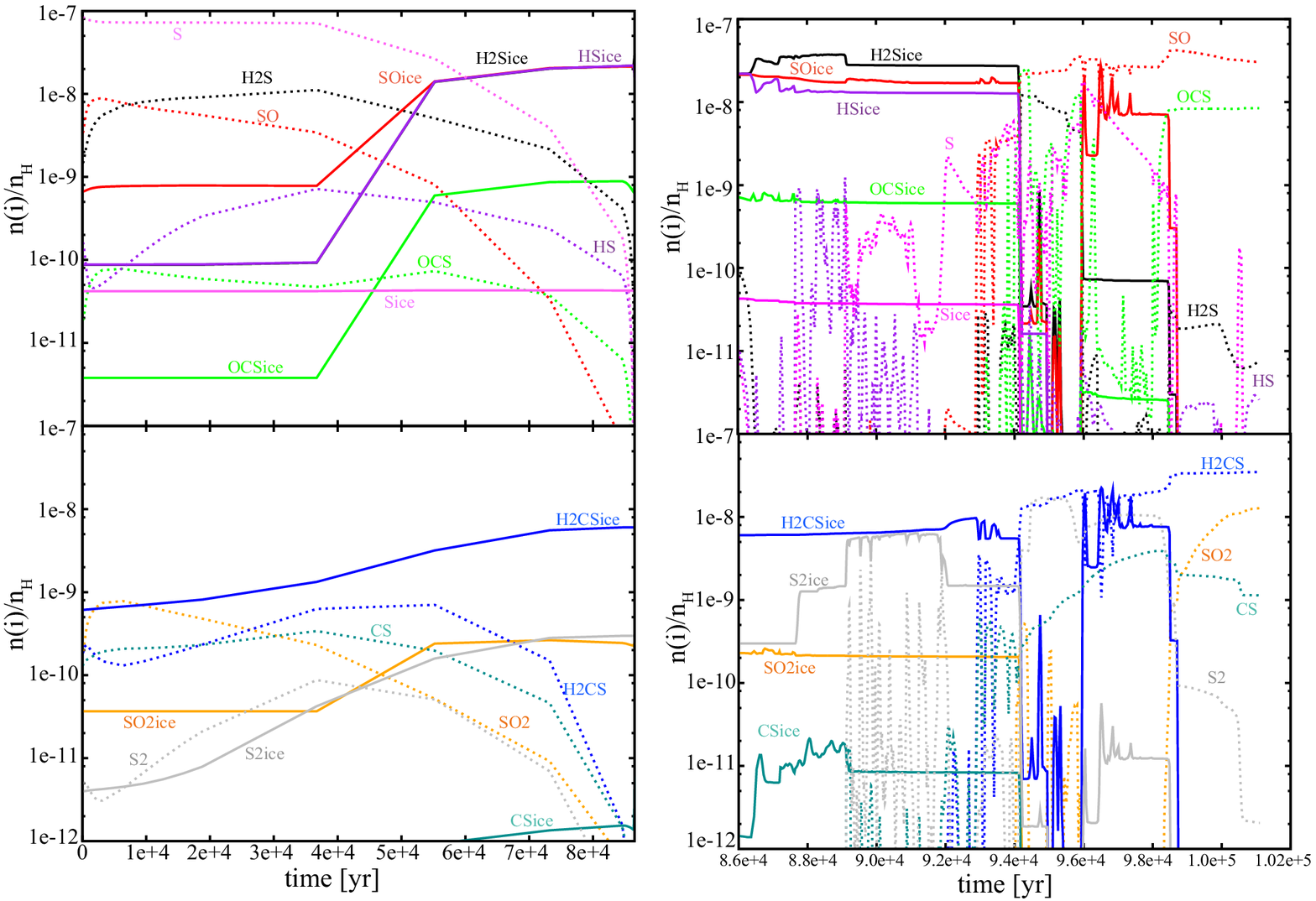}
\caption{cont.
}
\end{figure}

\begin{figure}
\epsscale{0.7}
\plotone{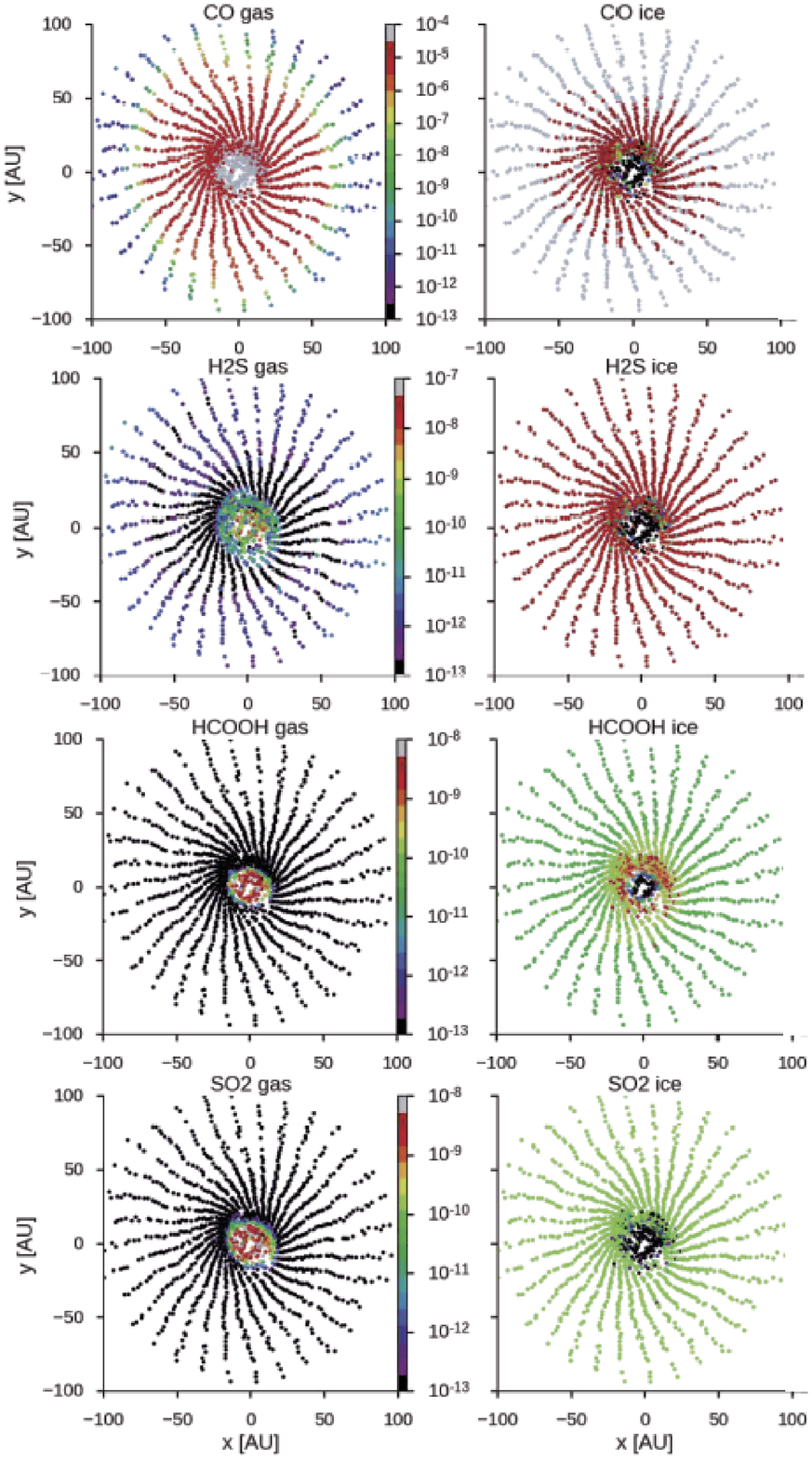}
\caption{Distribution of extracted SPH particles color coded by the molecular abundances at the final time step.
CO represents the molecules originate in the cloud core before the onset of collapse. H$_2$S also originates in the cloud, but is
destroyed after sublimation. HCOOH and SO$_2$ represent the molecules originate in the disk.
\label{dist_faceon}}
\end{figure}

\begin{figure}
\epsscale{0.7}
\plotone{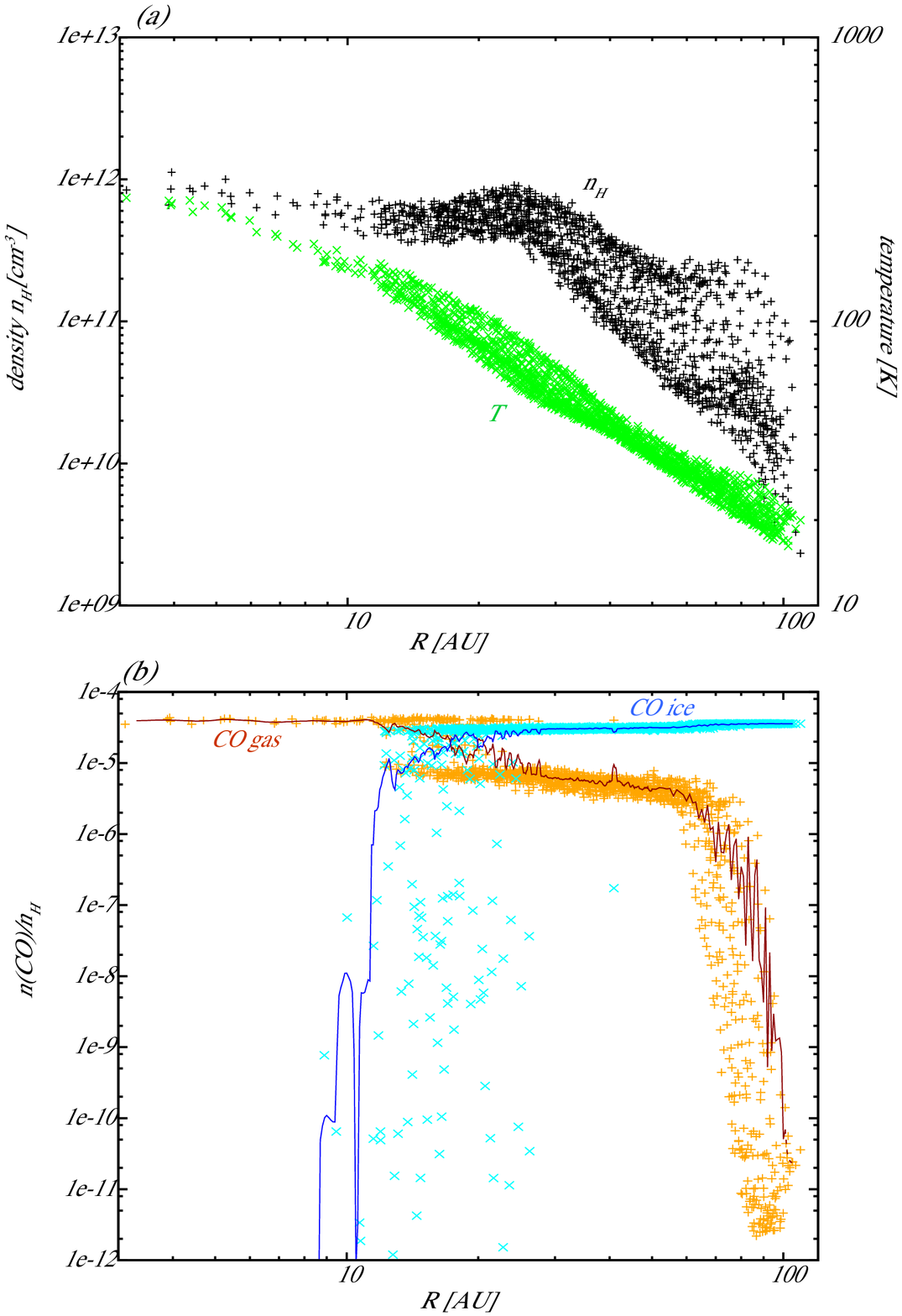}
\caption{Radial distribution of density and temperature (a) and relative abundance of CO to hydrogen nuclei in the gas phase and on grain
surfaces (b). The solid lines in panel (b) show the azimuthally averaged abundances.
\label{r_CO}}
\end{figure}

\begin{figure}
\epsscale{1.0}
\plotone{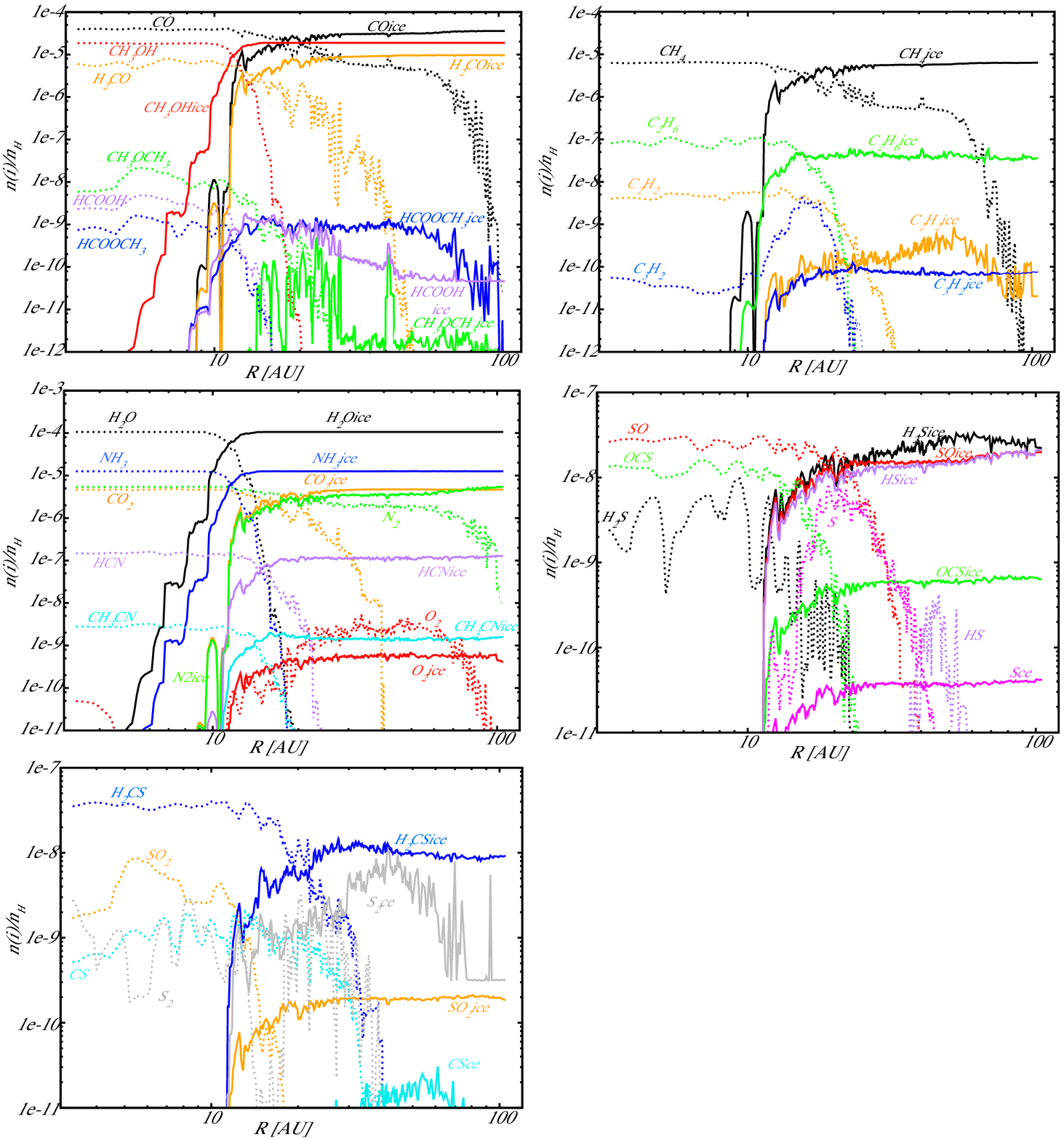}
\caption{Radial distributions of azimuthally averaged molecular abundances.
\label{radial_dist_abun}}
\end{figure}

\begin{figure}
\epsscale{1.0}
\plotone{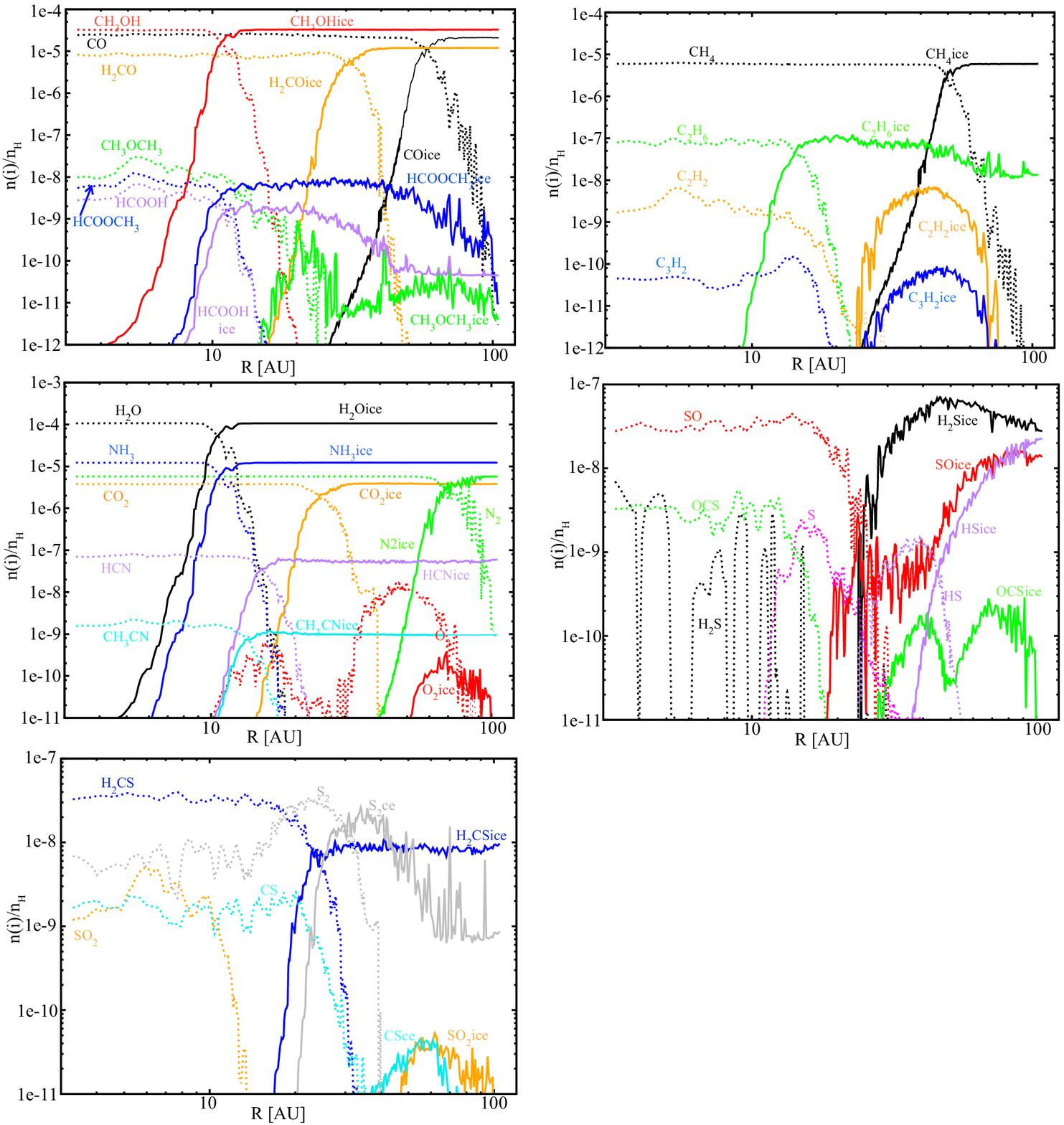}
\caption{Radial distribution of azimuthally averaged molecular abundances as in Figure \ref{radial_dist_abun} but for the two-phase model.
\label{radial_dist_2ph}}
\end{figure}

\begin{figure}
\epsscale{1.0}
\plotone{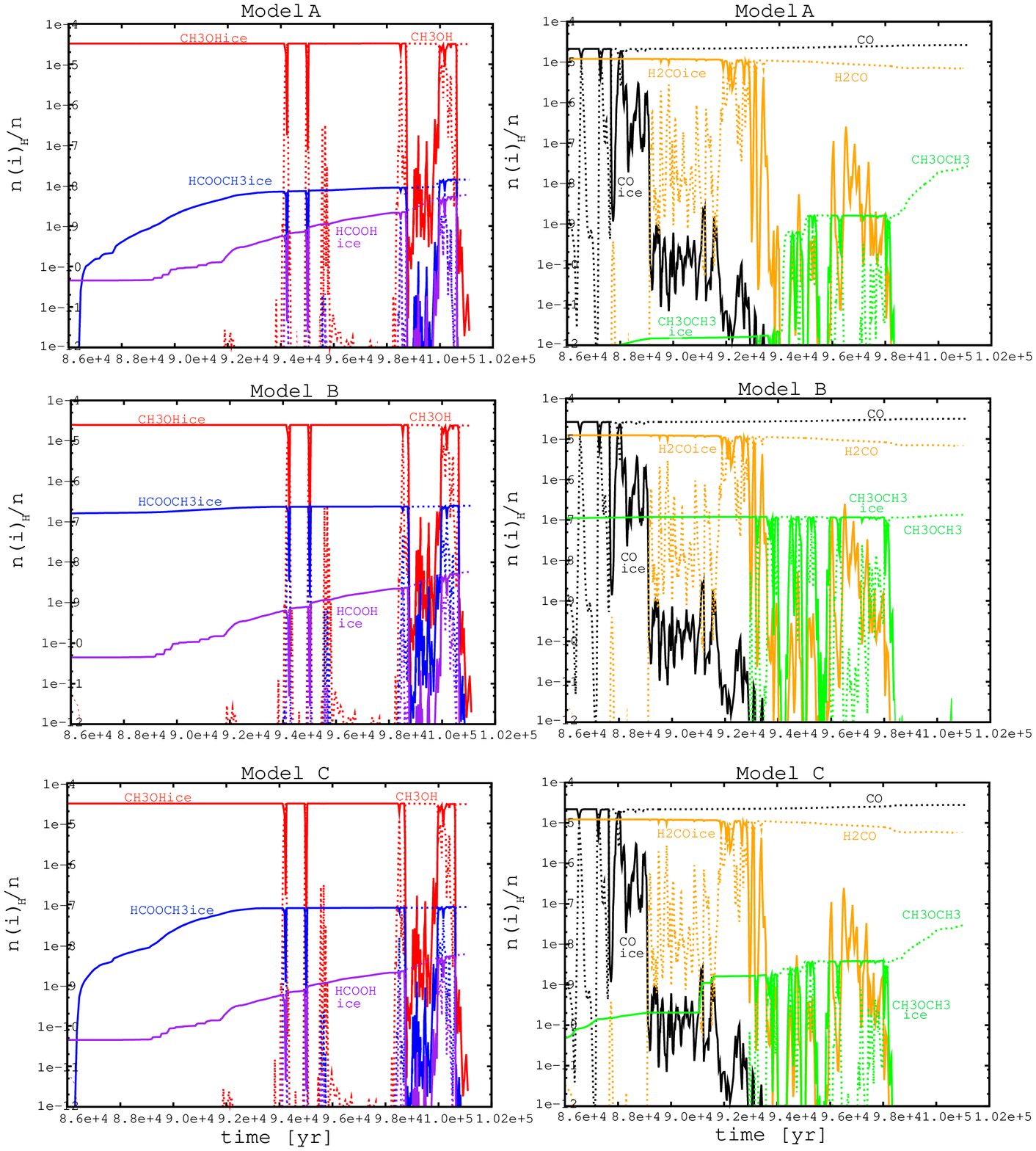}
\caption{ Temporal variation of molecular abundances obtained by the two-phase models in the SPH particle  along the trajectory shown in Figure \ref{traj1948}. The upper, middle, and the bottom panels show Model A, B, and C, respectively.
\label{COMs_test}}
\end{figure}

\end{document}